\documentclass[conference]{IEEEtran}
\IEEEoverridecommandlockouts

\usepackage{graphicx}
\usepackage{amssymb}
\usepackage{amsmath}
\usepackage{algorithm}
\usepackage{algorithmicx}
\usepackage{algpseudocode}
\usepackage{comment}
\usepackage{xspace}
\usepackage{float}
\usepackage{array}
\usepackage{subfig}
\usepackage{booktabs}

\def\BibTeX{{\rm B\kern-.05em{\sc i\kern-.025em b}\kern-.08em
    T\kern-.1667em\lower.7ex\hbox{E}\kern-.125emX}}

\begin{document}

\title{Neural Acceleration for Graph Partitioning
\thanks{This work was funded by the National Science Foundation grant \#2044633.
This work used Bridges2 at Pittsburgh Super Computing through allocation CCR180052 from the Advanced Cyberinfrastructure Coordination Ecosystem: Services \& Support (ACCESS) program, which is supported by National Science Foundation grants \#2138259, \#2138286, \#2138307, \#2137603, and \#2138296.}

}

\author{\IEEEauthorblockN{Joshua Dennis Booth}
\IEEEauthorblockA{\textit{Department of Computer Science} \\
\textit{University of Alabama in Huntsville}\\
Huntsville AL, USA \\
joshua.booth@uah.edu}
\and
\IEEEauthorblockN{Vishvam Patel}
\IEEEauthorblockA{\textit{Department of Computer Science} \\
\textit{University of Alabama in Huntsville}\\
Huntsville AL, UAH\\
email address }

}

\maketitle

\begin{abstract}
Graph Partitioning is a critical problem in numerous scientific and engineering domains including social network analysis, VLSI design, and many more. 
Spectral methods are known to produce quality partitions while minimizing edge cuts for a wide range of problems. 
However, the computational cost associated with the calculation of the Fiedler vector, an eigenvector associated with the second smallest eigenvalue of the graph Laplacian, remains a significant bottleneck due to memory issues and computational costs. 
In this paper, we present an accelerated approach to spectral bisection partitioning by replacing the traditional eigenvalue calculation with a simple artificial neural network model to approximate the Fiedler vector. 
We demonstrate that our approach achieves partitioning quality comparable to spectral bisection while significantly reducing the computational overhead, making it more scalable and efficient for large-scale problems.
\end{abstract}






\section{Introduction}
\label{sec:intro}
Graph partitioning is a fundamental problem used by applications in various fields,
including scientific computing, VLSI design, and parallel computing~\cite{vlsi, spec_par_map}. 
While there exist multiple variations of this problem, we focus on the following common one.
This variation involves dividing a graph into two approximately equal-sized sub-graphs, also known as Graph Bisection (GB), while minimizing the number of edges that connect them.
This general problem is NP-hard.
However, there exists multiple heuristic method (e.g., spectral~\cite{spec_part} and multilevel~\cite{mlti_lvl_spec_part, metis} methods) that are commonly utilized.
One common method for partitioning graphs is the spectral bisection (SB) method, as it is known to generate reasonably good partitions compared to other methods such as geometric methods. 
However, SB is computationally expensive because it requires to calculate the Fiedler vector~\cite{fied}, which is the eigenvector associated with the smallest non-trivial eigenvalue.
While SB is effective for smaller or denser graphs, they often struggle with the challenges posed by large, sparse graphs, where maintaining a balance between computational efficiency and partition quality becomes increasingly difficult. 
In this work, we explore the use of neural acceleration to approximate the Fiedler vector in order to automate this process for scientific application users and better utilize the heterogeneous computing environments (i.e., CPU-GPU) common in high-performance computing.

Neural Acceleration~\cite{nn1,nn2} replaces key computationally expensive segments of an algorithm with cheap and simple artificial neural networks. 
The idea is aimed at modern workloads on heterogeneous systems to improve energy usage and
speed up execution time. 
These systems commonly contain GPU accelerators optimized for dense tensor computations used in training and evaluating neural networks. 
These neural networks are small enough to be easily trained and executed at compile time or in parallel to the application. 

The following workflow is used to apply neural acceleration. 
The programmer identifies computationally expensive functions that may not suffer from approximate solutions. 
To incorporate it at compile time, the programmer provides a small amount of data about the function such as expected inputs, outputs, and computational flow. 
A cheap neural network is then trained on the GPU. 
At the time of execution, the neural network is utilized on the GPU instead of the original function.

The spectral method is the ideal choice for neural acceleration over other GB methods because of its mathematical foundation.
Unlike geometric or multilevel methods, SB transforms the GB problem into a continuous optimization problem. 
SB leverages the eigenvector associated with the second smallest non-trivial eigenvalue to partition the graph.
To calculate the eigenvalue and eigenvector, SP performs an eigen decomposition of the Laplacian matrix to get the eigenvalues and eigenvectors.
The eigen decomposition is considered an optimization problem; hence, so is SB.
Neural networks are known to approximate continuous functions, as seen in image processing, time series forecasting, etc.
We present a neural acceleration method capable of predicting the Fiedler vector while maintaining quality and performance.

This paper first provides the formal definition of this problem along with common solutions in Section 2.
Section 3 presents the neural acceleration model used for our graph partitioning problem.
Section 4 provides our experimental setup, and Section 5 presents our evaluation in terms of quality and performance.

\section{Problem Definition and Background}
\label{sec:background}
In this section, we formally define the graph partitioning problem and provide a high-level overview of the partitioning methods.

\subsection{Problem Definition}
\label{subsec:part_def}
Let \( G = (V, E) \) be an undirected and unweighted graph. 
The objective is to partition the vertex set \(V\) into two disjoint sets, \(V_1\) and \(V_2\), of roughly equal size, while minimizing the number of edges between them. 
Commonly referred to as Graph Bisection (GB).
The number of edges that connect the vertices in \( V_1 \) and \( V_2 \) are known as the edge cut. 
The edge cut \( EC \) is defined as follows:
\begin{equation}
\label{eq:edge-cut}
    EC = \sum_{\substack{i\in V_1\\ j\in V_2}} e_{i,j}~.
\end{equation}
Therefore, the objective function becomes
\begin{equation}
    \min \sum_{\substack{i\in V_1\\ j\in V_2}} e_{i,j} ~\text{s.t.}~ \lvert V_1 \rvert \approx \lvert V_2 \rvert ~.
\end{equation}
Many graph partitioning algorithms aim to minimize this edge cut, as it represents the total number of edges between the two partitions. 
However, some algorithms focus on minimizing the edge cut ratio, which normalizes the edge cut by the total number of edges in the graph. 
The edge cut ratio \( ECR \) is defined as:
\begin{equation}
\label{eq:edge-cut-ratio}
    ECR = \frac{EC}{|E|}
\end{equation} 
where \(|E|\) is the total number of edges.
Inherently, improving one improves the other and vice versa.

In this work, we consider the unweighted, bisection version of graph partitioning, but the definition can be extended to include weighted graphs and $ k $-way partitioning for more general cases.

\subsection{Background and Related Work}
In this section, we will provide an overview of the current state-of-the-art methods in graph partitioning.
\subsubsection{Spectral Methods}
\label{subsubsec:spec_mthds}
Spectral Partitioning (SP) divides a graph into two parts by leveraging the eigenvectors of its Laplacian matrix, also known as Spectral Bisection (SB).  
The Laplacian matrix of a graph, denoted by $L$, is given by $L = D - A$, where $A$ is the adjacency matrix of the graph, and $D$ is the diagonal degree matrix. Such that, each element $A[i,j] = 1$ represents an edge between vertex $i$ and $j$, while each diagonal entry $D[i,i]$ represents the degree (number of adjacent vertices) of vertex $i$. Therefore, the graph Laplacian matrix $L$ is represented element-wise as follows:
\begin{equation}
    L[i,j] = 
        \begin{cases}
             deg(v_i) & \text{if } i = j,\\
             -1 & \text{if } i \neq j \text{ and } A_{i,j} = 1,\\
             0 & \text{if } i \neq j \text{ and } A_{i,j} = 0.
        \end{cases}
\end{equation}
The graph Laplacian is symmetric positive semi-definite (i.e., eigenvalues $\lambda_i \geq 0$ for all $i$). 
It is a key component in SP because of its eigenvalues and eigenvectors, which reveals important details about the structure of the graph. 
Given that the graph $G$ is connected, the eigenvalues of $L$ are non-negative. 
The smallest eigenvalue $\lambda_1$ is zero and is considered trivial. 
Hence, we extract the second smallest non-trivial eigenvalue $\lambda_2$ (known as the algebraic connectivity of a graph) and its corresponding eigenvector, the Fiedler vector~\cite{fied}. 
Find the median of the values in the Fiedler vector. 
Assign the vertices whose corresponding values in the Fiedler vector are less than the median to one partition and the vertices with values greater than or equal to the median to the other partition. 
You can further apply SB recursively to find more partitions of the graph, known as Recursive Spectral Bisection (RSB).

A major drawback of SP is the computational complexity associated with eigenvalue calculation and the memory requirement to store the graph. 
As the size of the graph grows, it is impractical to perform spectral partitioning at the original size as the whole graph will not fit in memory.
In particular, this is due to the fact that even though the Laplacian is sparse (i.e., contains mostly zeros) that can be stored in compressed format, the eigenvalue calculation is dense (i.e., the memory of $O(\vert V \vert^2$)).
Multilevel partitioning methods were developed to overcome the memory problem. 
More on that in Section~\ref{subsec:mlvl_mthds}.
Recently, GPUs have been leveraged for spectral partitioning using tools like SpyhNX~\cite{sphynx}, which implements spectral partitioning on GPUs. 
This is particularly relevant to our work, as it enables spectral methods to handle large graphs more efficiently by parallelizing the eigenvalue computation across many cores. 
By accelerating this process, SpyhNX can address one of the major drawbacks of traditional spectral methods, which is their computational intensity, especially when applied to large graphs.

\subsubsection{Geometric Methods}
\label{subsec:geo_mthds}
Geometric Partitioning (GP) utilizes the geometric or spatial coordinates associated with the vertices to divide a graph into two or more parts. 
These methods are particularly effective when the graph originates from physical simulations or meshes, where the vertices represent points in space. 
One of the simplest GP algorithms is Recursive Coordinate Bisection (RCB)~\cite{rcb}, which partitions the graph by selecting a coordinate axis and dividing the vertices based on the median value along that axis. 
Vertices on one side of the median are assigned to one partition, while those on the other side are placed in the second partition. 
This process is repeated recursively for each partition until the desired number of parts is achieved. 
RCB is similar to RSB, except that one utilizes physical coordinates and the other uses an eigenvector.
More advanced methods would include Geometric Mesh Partitiong (GMP)~\cite{gmp} and Scalapart~\cite{scala_part} for parallel implementation.

A major limitation of GP is their dependence on the availability of graph coordinates.
However, many real-world graphs, such as those from social networks, do not have coordinates associated with them.
To address this, various methods have been developed to embed graphs and generate their coordinates.
The force-based embedding algorithm by Hu~\cite{force_embed}, originally developed for graph drawing, is adapted to generate coordinates in methods like Scalapart. 
This algorithm calculates the N-body forces between vertices using the Barnes-hut approximation to embed the graph, achieving a time complexity of $O(n\log n)$.
Although this approach is effective in producing high-quality embeddings, it is important to note that not all graphs can be embedded in space.
The overhead of generating an embedding (with an uncertainty of quality) combined with the cost of the GP method outweighs the benefits.

\subsubsection{Hierarchical Methods}
\label{subsec:mlvl_mthds}
Hierarchical or multilevel partitioning methods are fundamental in graph partitioning, where a graph is coarsened into a smaller representation, partitioned, and then refined as it is uncoarsened back to its original size.
Multilevel Spectral Bisection~\cite{mlti_lvl_spec_part} is one of the prominent methods. 
This method first reduces the size of the graph by collapsing adjacent vertices and edges to form a series of coarser graphs. 
Spectral partitioning is then applied to the coarsest graph and progressively refined as it is uncoarsened.
Another state-of-the-art method is METIS~\cite{metis}.
The graph is coarsened using the heavy-edge matching (HEM) algorithm, which collapses adjacent vertices with the heaviest edge.
Next, it performs multiple combinations of partitions by randomly assigning vertices at the coarsest level and chooses the partition with the lowest edge cut.
As it uncoarsens the graph, partition refinement (e.g., Fiduccia-Mattheyses~\cite{fm}) is applied near the partition boundary.
METIS produces high-quality partitions while being extremely efficient.
However, if the initial partition is not chosen correctly, which is highly likely because of its randomness, it can produce suboptimal partitions with a high edge cut.
We consider the simple version of FM in our approach.

\subsubsection{Deep Learning Approaches}
\label{subsubsec:gap}

The Generalized Approximate Partitioning (GAP) framework~\cite{gap_part} applies deep learning to graph partitioning.
GAP divides the framework into two modules: embedding and partitioning. 
The embedding module begins by coarsening the graph down to two nodes and applies a GraphSAGE convolution (SAGEConv) to create a feature representation, starting with a 2x2 identity matrix at the coarsest level. 
As the graph is interpolated back to finer levels, shared SAGEConv layers are applied, followed by hyperbolic tangent activation (tanh). 
At the finest level, linear layers followed by QR factorization are used to compute the two smallest eigenvalues and their eigenvectors. 
The partitioning module similarly coarsens the graph but uses three distinct sets of SAGEConv layers applied before, during, and after coarsening, each followed by tanh activation. 
Again, at the finest level, linear layers followed by a softmax function predict the partition probabilities.
The embedding and partitioning modules minimize the Eigenvector Residual and Expected Normalized Cut objective function, respectively.
Together, these modules generate spectral embeddings and predict partitions in an unsupervised fashion.

GAP framework produces partition quality comparable to state-of-the-art methods like Metis and Scotch, however, there are a few limitations to the framework. 
Firstly, the multilevel coarsening and interpolation steps, coupled with multiple GraphSAGE convolution layers, significantly increase the complexity of the algorithm. 
Furthermore, GAP is highly computationally intensive due to QR factorization, which requires $O(n^3)$ FLOPS to extract eigenvalues and eigenvectors in the embedding module. 
This factorization, along with the calculations of Rayleigh quotient (requires $O(n^2 + C)$ FLOPS) for eigenvalue residuals, demands heavy computational resources.
From a deep learning perspective, GAP has a rather small number of trainable parameters ($\sim10,000$). 
However, the training time for the embedding and partitioning module exceeds 10 hours, where each module was trained for 120 epochs.
The high training time is a direct reflection of the computational complexity of the algorithm.

\subsubsection{Partitioning Refinement}
\label{subsubsec:refinement}

In graph partitioning, refinement is a crucial step in improving partition quality after the initial partition is obtained.
The quality of initial partitions is often based on coarsening-based heuristics.
Frameworks such as Metis, Sphynx, and Scotch~\cite{scotch} typically follow this approach.
However, due to the randomness of the coarsening techniques, the resulting partitions may be suboptimal leading to imbalances in partition sizes and unnecessary edge cuts between partitions.
The refinement step is designed to address these issues. 
After an initial partition is obtained, the refinement techniques work by adjusting and improving the partition assignments to achieve better balance and reduce edge cuts.

The two most commonly used refinement techniques are Fiduccia-Mattheyses (FM)~\cite{fm} and Kernighan-Lin (KL)~\cite{kl}.
Both techniques improve the partition by greedily moving nodes from one partition to another based on a gain function. 
While they operate similarly, the only difference is that FM moves one node at a time, whereas KL moves a pair of nodes.
KL requires to find a pair of nodes in different partitions to be swapped, which further minimizes the edge cut, resulting in a runtime of $O(n^2\log n)$, while FM performs in $O(n)$.
Although KL is known to produce better quality cuts, FM is preferred due to its faster performance.
\begin{figure*}[tbh]
    \centering
      \includegraphics[width=.65\textwidth]{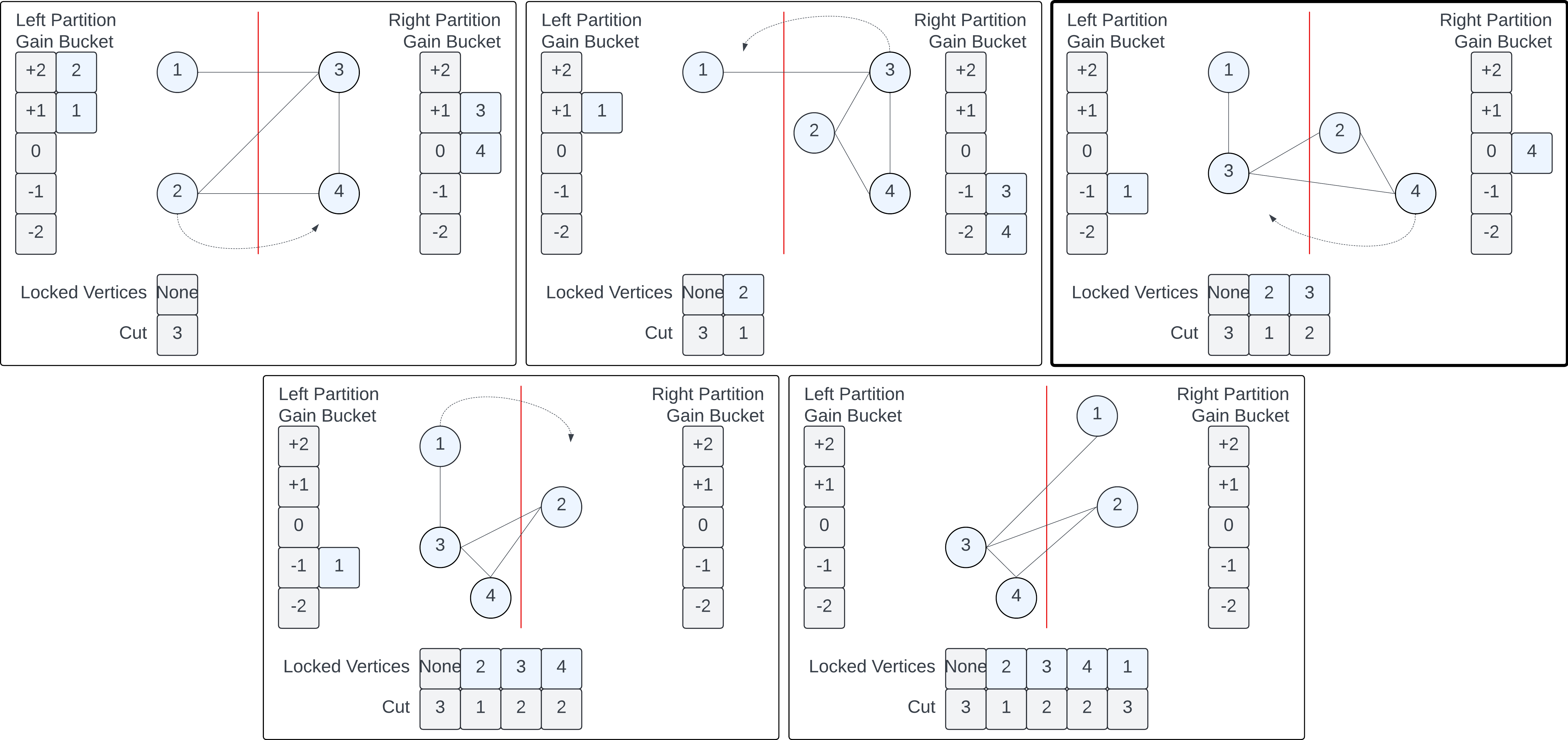}
    \caption{A complete pass of Fiduccia-Mattheyses Algorithm. The best edge cut is highlighted.}
    \label{fig:fmOnePass}
\end{figure*} 
FM operates by computing a gain function for each node (i.e., essentially measuring how beneficial it would be to move the node from one partition to another) and assigns them to their respective gain buckets.
The algorithm iteratively moves the nodes with the highest gain while maintaining balance until no further improvements can be made. 
After each node has been moved, it backtracks to the best move with the desired balance. 
An illustration of the FM algorithm is shown in Figure \ref{fig:fmOnePass}.

Many state-of-the-art graph partitioning frameworks implement variations of the FM algorithm. 
METIS, for instance, focuses on refinement near the partition boundaries rather than across the entire graph. 
This localized refinement ensures computational efficiency while still allowing for significant improvements in partition quality. 
By restricting adjustments to nodes near the boundary, METIS minimizes computational overhead, speeding up the refinement process while still addressing the most critical areas for improvement. 
In our approach, we consider the simpler version of FM.

\section{Neural Acceleration Framework For Graph Partitioning}
\label{sec:alg}

Artificial Neural Networks (ANN) have shown an exceptional ability to capture complex patterns in big data, leading to their vast application in numerous fields such as image processing, natural language processing, and autonomous systems. 
The trend is to build large, deep neural networks that require a tremendous amount of data to train. 
Due to the high computational complexity of such models, the time required to train is often significantly high, which presents a challenge when time and resources are limited. 
The GAP framework (mentioned in section 2.2.4) follows this trend, where they apply GraphSAGE~\cite{graphsage} convolution in a multilevel fashion along with dense layers to obtain the partition of a graph. 
Due to graph partitioning being a NP-Hard problem, it is extremely difficult to model it using ANNs without requiring additional computation overhead.
For such a reason, GAP adapted two custom objective functions to achieve state-of-the-art quality edge cuts, which added more complexity to the framework.
However, our goal is to accelerate the key computation segment of an expensive algorithm using ANN.
In spectral partitioning, the calculation of the Fiedler vector is the most expensive process.
Additionally, finding the Fiedler vector is an optimization problem, whereas predicting the partition directly is difficult to model using ANNs.
Thus, it is crucial to build a simple model to approximate the Fiedler vector capable of being trained on the GPU at compile time. 
We achieve this task using a simple one-hidden layer dense neural network in a supervised manner.

\begin{figure}[tbh]
    \centering
    \includegraphics[width=.37\textwidth]{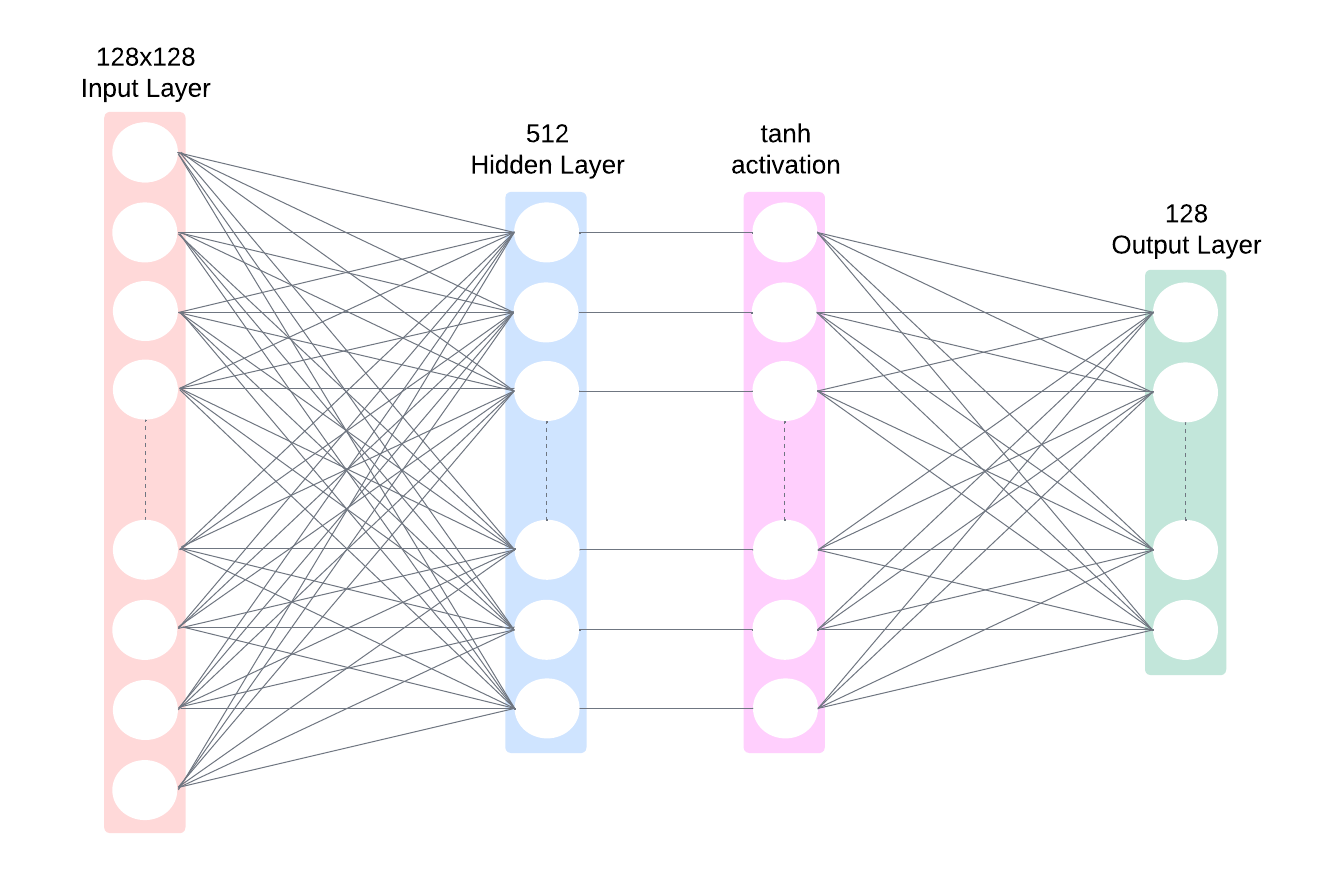}
    \caption{Model Architecture}
    \label{fig:nn-arc}
\end{figure}

The choice of neural network architecture is critical.
Since we are working with graph data, graph neural networks (GNNs) seem like a natural choice, as they have gained significant attention in recent years for their ability to capture graph data accurately. 
However, GNNs rely heavily on the availability of node features to make accurate predictions, and they often struggle to learn effectively when only the graph structure is available. 
Additionally, graph convolution operations in GNN architectures such as GCN~\cite{gcn}, GIN~\cite{gin}, and GAT~\cite{gat} are computationally expensive, which further reduces their suitability for our needs.
On the other hand, GraphSage has proven efficient due to their simple and fast message-passing aggregation; it requires node features to learn the data.
Since graphs do not have node features associated with them, hence no message passing, we can rely on linear transformations only.
For this reason, we opt for a simpler approach by using dense layers to construct our model.

Our model is a one-hidden layer dense neural network. 
It is defined as follows. 
The input is a flattened graph Laplacian matrix of size 128x128. 
The hidden layer has 512 units with bias, followed by tanh activation. 
The output layer has 128 units without bias, which predicts the Fiedler vector. 
The model has 8,454,656 trainable parameters. 
Figure~\ref{fig:nn-arc} provides a visual representation of our model. 
The problem size of 128 vertices provides a reasonable balance between the model's complexity while maintaining complex and diverse connectivity patterns in graphs. 
Graphs with smaller size could potentially lead to trivial Fiedler vectors. 
Graph Laplacian is used as input to model the problem as closely to spectral partitioning as possible. 
Furthermore, graph Laplacian matrix represents important spectral properties including number of connected components and community structure.
A hidden layer with 512 units provides sufficient capacity to learn meaningful representations of the graph Laplacian. 
The large number of parameters in this model could potentially be an issue. 
Although preprocessing the input data to extract fewer, more meaningful features is an option, it would involve significantly more preprocessing, which does not align with our goal of keeping the process simple. 
On the other hand, GPUs are optimized for dense tensor multiplications, which allows us to efficiently train the model despite its relatively large number of parameters.
With this model, we aim to strike the balance of achieving simplicity without sacrificing quality.

\begin{algorithm}
\caption{Local Neighborhood Filter}\label{algo:sw}
\begin{algorithmic}[1]
    \footnotesize
    \Function{LocalFilter}{$A, \text{targetSize}$}
        \State $n \gets size(A)$
        \State $M \gets zeros(targetSize, targetSize)$
        \State $highStep \gets ceil(n / targetSize)$
        \State $lowStep \gets floor(n / targetSize)$
        \State $boundary \gets targetSize - (targetSize * highStep - n)$
        \For{$i \gets 1$ \textbf{to} $targetSize$}
            \If{$i \leq boundary$}
                \State $X \gets (i-1) * highStep + 1 \text{ to } i * highStep$
            \Else
                \State $X \gets (i-1) * lowStep + 1 \text{ to } i * lowStep$
                \State $X \gets X + boundary$
            \EndIf
            \For{$j \gets 1$ \textbf{to} $targetSize$}
                \If{$j \leq boundary$}
                    \State $Y \gets (j-1) * highStep + 1 \text{ to } j * highStep$
                \Else
                    \State $Y \gets (j-1) * lowStep + 1 \text{ to } j * lowStep$
                    \State $Y \gets Y + boundary$
            \EndIf
            \State $tile \gets A[X, Y]$
            \State $M[i, j] \gets mean(tile)$
            \EndFor
        \EndFor
        \State \Return $M$
    \EndFunction
\end{algorithmic}
\end{algorithm}

We presented a neural network model for a specific size of vertices. 
Here, we will provide a two-level framework to scale our approach to bigger graphs. 
Most multilevel methods use coarsening heuristics such as Heavy-Edge Matching (HEM) to obtain smaller version of the graph. 
However, our coarsening method is based on the local neighborhood averaging or voting filter used in image processing~\cite{woods}. 
Instead of operating on the graph structure itself, our method operates on the adjacency matrix.
Additonally, a model of this size can fit comfortably on a mid-range discrete GPU.
\begin{figure*}[tbh]
    \centering
    \includegraphics[width=.65\textwidth]{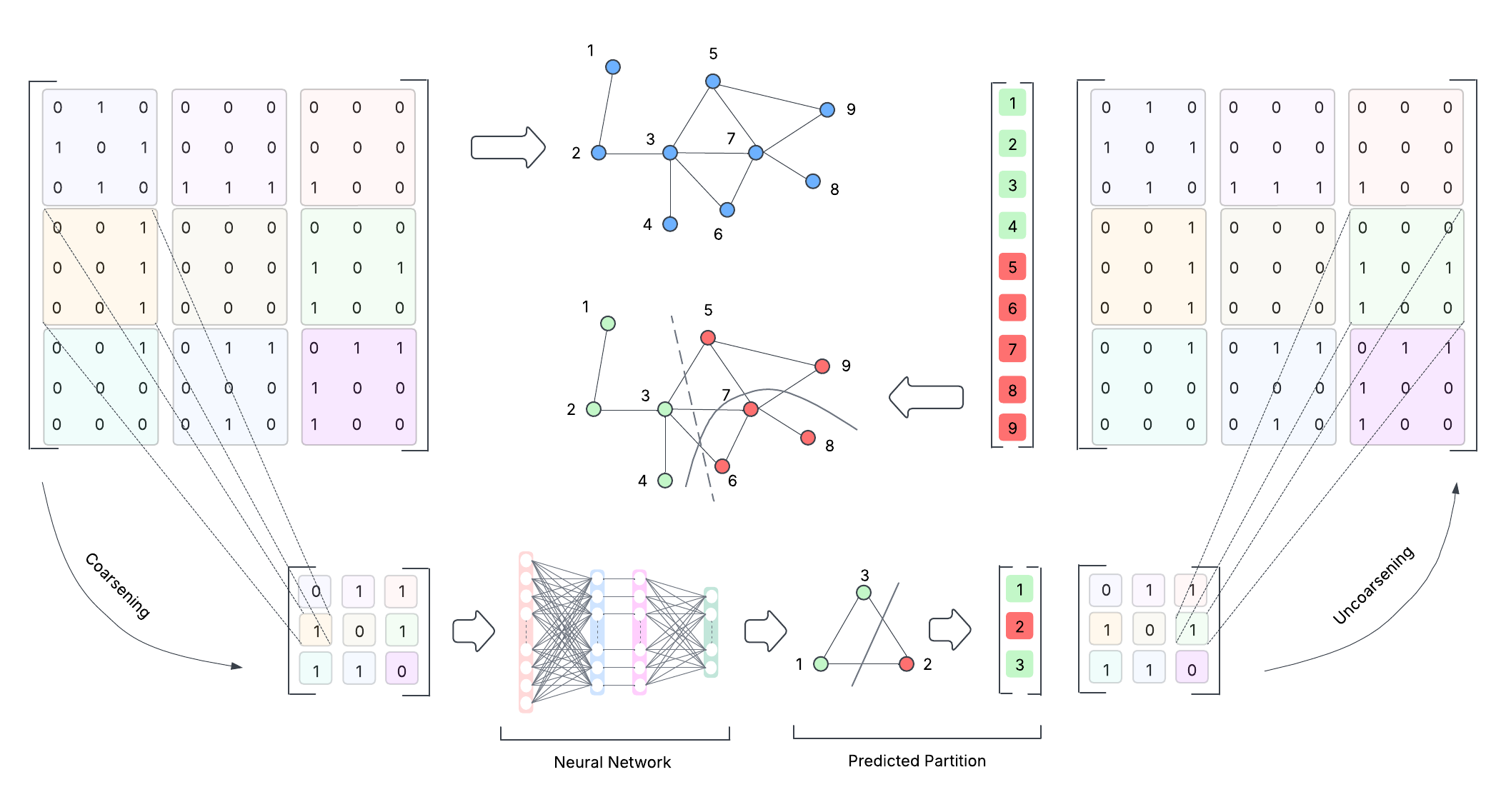}
    \caption{Our Multilevel Scalable Workflow}
    \label{fig:EvalWorkflow}
\end{figure*}
The idea is as follows: calculate a filter size based on the target size, slide the filter over the adjacency matrix, find the average of the filter, assign it as an edge if it is greater than 0, and store the mappings. 
Our coarsening approach is provided in Algorithm~\ref{algo:sw}. 
The algorithm's runtime complexity is $O(n^2)$ in the dense case, however, it improves significantly when stored in sparse format. 
The runtime complexity becomes $O(m^2 + nnz)$, where $m$ is the target size and $nnz$ is the number of non-zeros. 
Additionally, this algorithm could be implemented efficiently in parallel on a GPU.
After coarsening the matrix, the method  applies our neural network model followed by FM refinement, and interpolates back to the original graph. 
The workflow is describe in Figure~\ref{fig:EvalWorkflow}. 

\section{Experimental Setup}
\label{sec:expsetup}

\subsection{Dataset}
\label{subsec:dataset}

\begin{figure}[tbh]
    \centering
    \subfloat[Graph 19]{
        \includegraphics[width=.30\textwidth, height=.2\textheight]{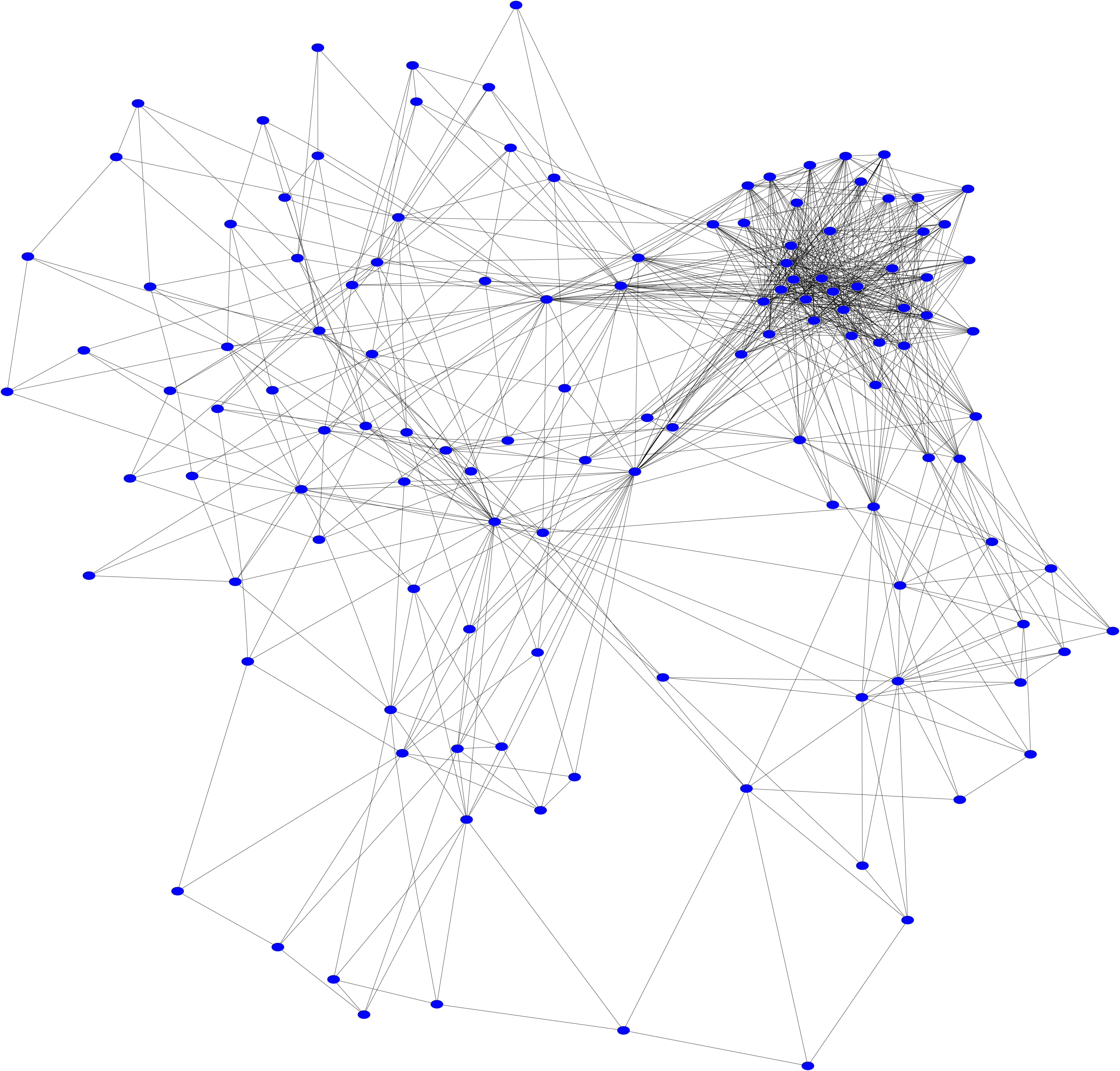}
    }\\
    \subfloat[Graph 45]{
        \includegraphics[width=.30\textwidth, height=.1\textheight]{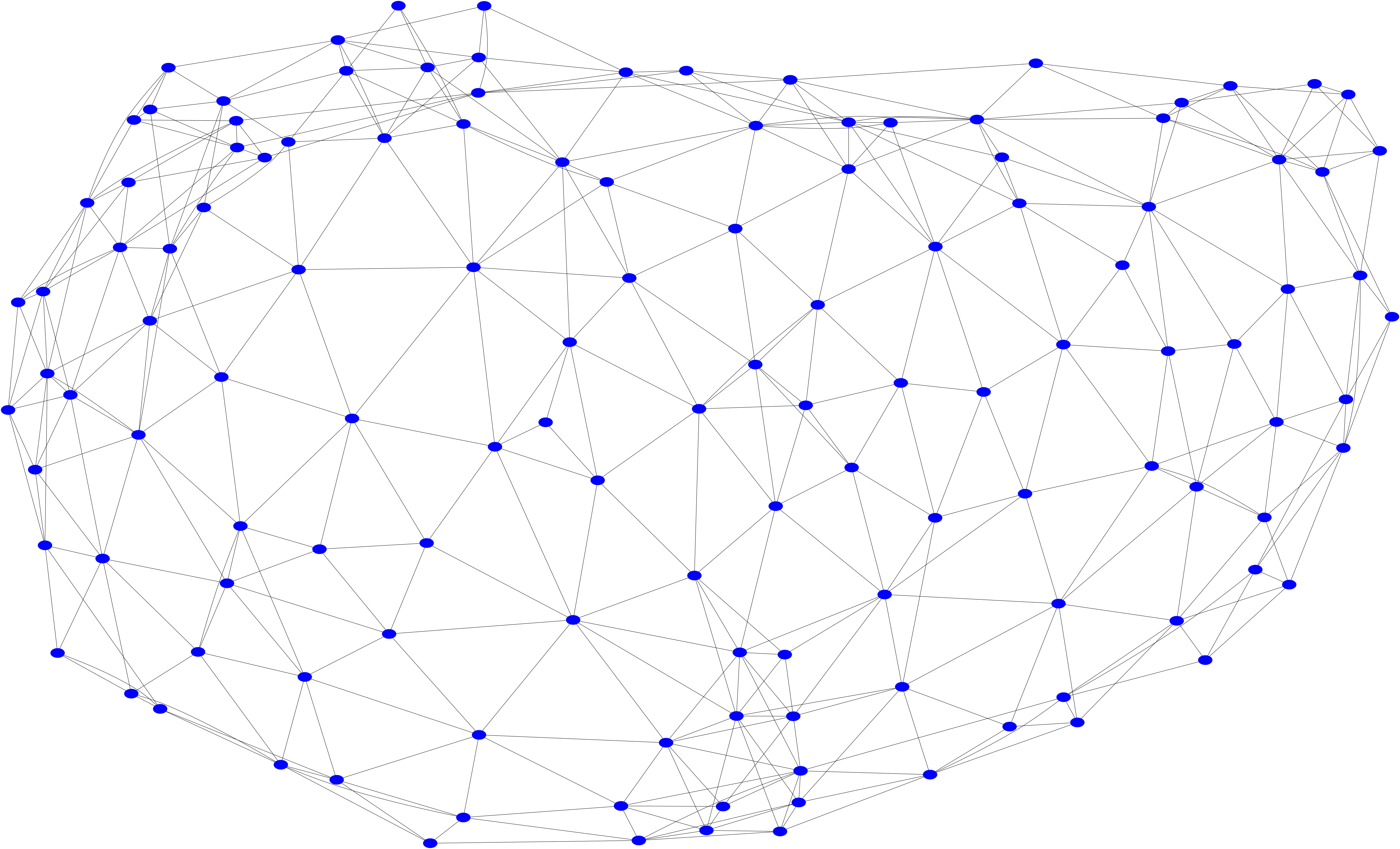}
    }\\
    \subfloat[Graph 106]{
        \includegraphics[width=.25\textwidth, height=.1\textheight]{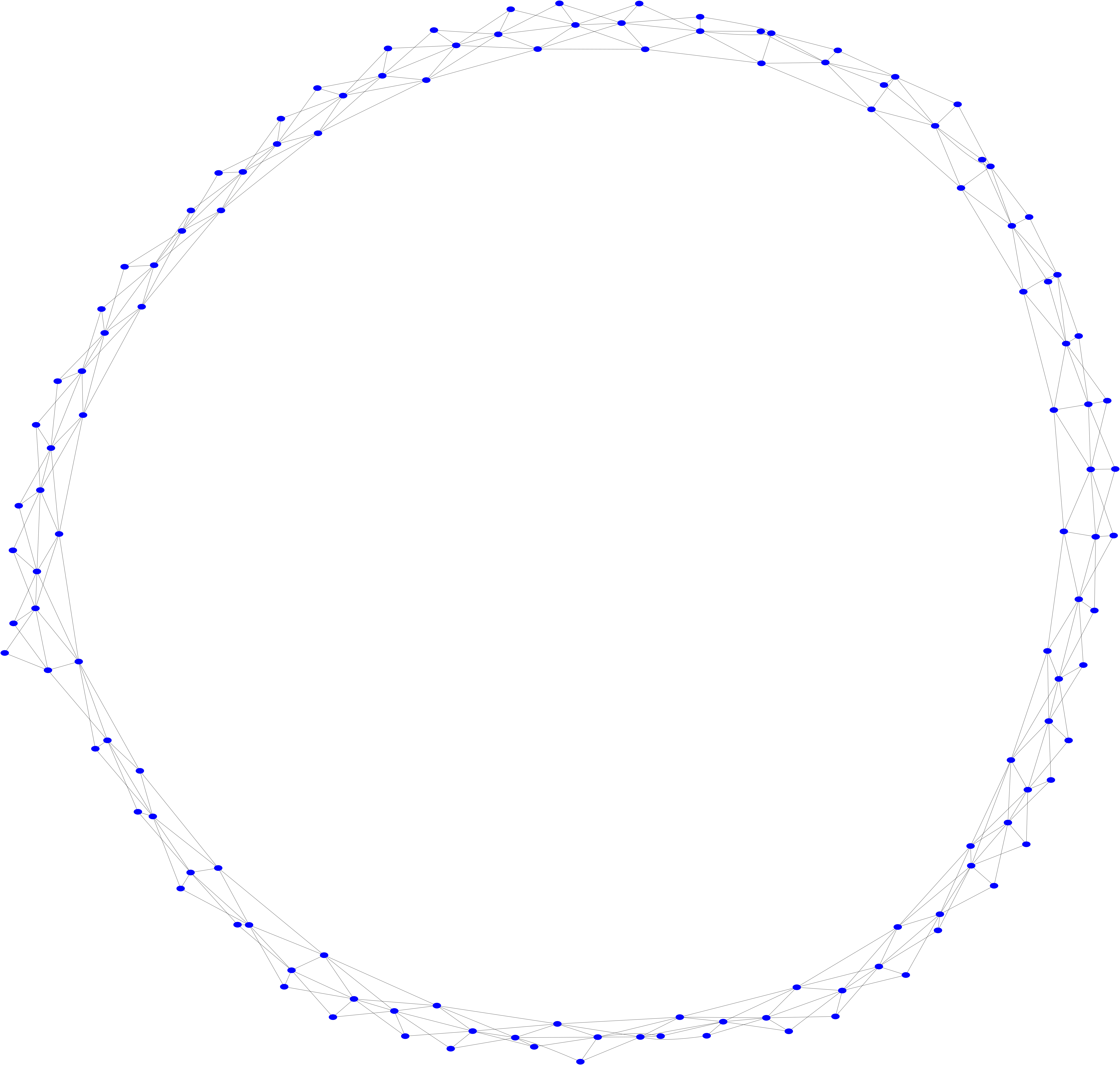}
    }\\
    \subfloat[Graph 8]{
        \includegraphics[width=.25\textwidth, height=.2\textheight]{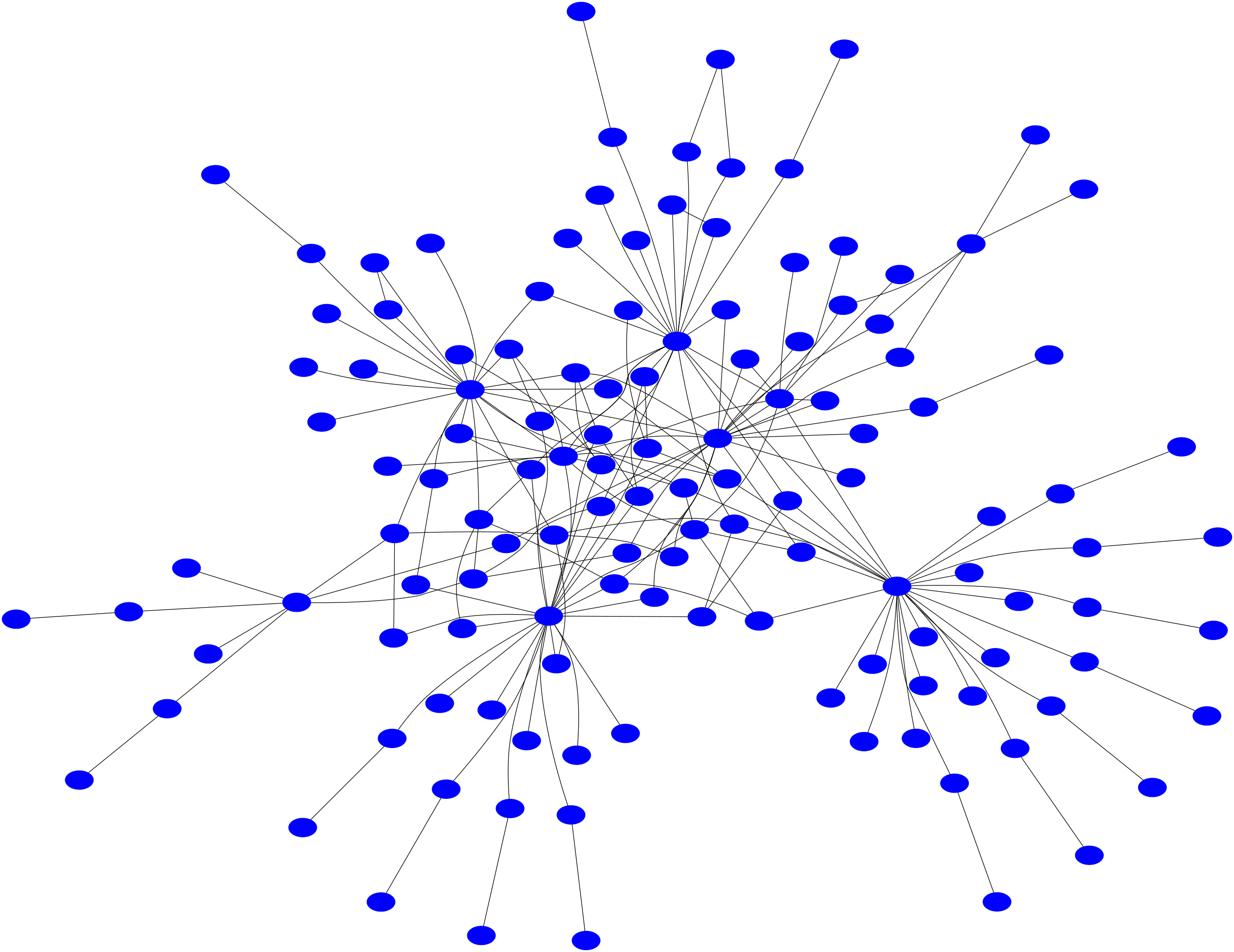}
    }
    \caption{Randomly Selected Test Graphs' Embeddings}   
    \label{fig:embeds}
\end{figure}

\begin{table}[t]
    \centering
    \begin{tabular}{c c c c c}
        \hline
        \multicolumn{2}{c}{} & \multicolumn{3}{c}{Degree}\\
        \cmidrule(lr){3-5}
        Graph & Edges & Avg & Max & Min \\
        \hline
        4   & 423  & 6.61   & 9   & 3 \\
        8   & 192  & 3.00   & 26  & 1 \\ 
        19  & 833  & 13.02  & 42  & 3 \\
        34  & 2602 & 40.66  & 127 & 8 \\
        45  & 420  & 6.56   & 11  & 3 \\
        66  & 7120 & 111.25 & 119 & 101 \\
        88  & 127  & 1.98   & 2   & 1 \\
        106 & 323  & 5.05   & 8   & 2 \\
        109 & 363  & 5.67   & 6   & 3 \\
        115 & 198  & 3.09   & 8   & 1 \\
        \hline
    \end{tabular}
    \caption{Characteristics of randomly selected graphs from the test dataset.}
    \label{tab:tdata_info}
\end{table}
We conduct two levels of experiments with each using a different test set.
The first level of experiments utilizes a small set of graphs for both training and testing of our neural network.
The objective is to have graphs with 128 nodes that can be used to train and test our inexpensive neural network models.
However, because of the lack of sufficient real-world data at this specific size, we generate data for our experiments by scaling other inputs. 
This set is generated by scaling down graphs and graph representations of sparse matrices from the SuiteSparse Matrix Collection~\cite{suitesparse}.
We select all undirected graphs with more than 512 and less than or equal to 1,000,000 nodes. 
To match our target size, larger matrices are scaled down using a $n$ x $n$ local neighborhood voting filter similar to image resizing with a filter~\cite{woods}. 
In order to enrich the dataset and introduce patterns that would improve model training, we apply different sparse matrix reordering techniques to the adjacency matrix representation of the graph.
Specifically, we use three popular ordering methods: Reverse Cuthill-McKee (RCM)~\cite{rcm}, Nested Dissection (ND)~\cite{nd}, and Approximate Minimum Degree (AMD)~\cite{amd}. 
Furthermore, scaled down graphs that resulted in disconnected graphs are removed.
The final dataset is divided into training and testing sets, with 1,156 graphs used for training and 126 for testing.
For reporting and visual purposes, we randomly select a set of graphs from the test dataset. 
The characteristics of these graphs are presented in Table~\ref{tab:tdata_info}, along with a visual embedding of a few graphs in Figure~\ref{fig:embeds}.

\begin{table}[tbh]
\footnotesize
    \centering
    \begin{tabular}{ccccccc}
        \hline
        \multicolumn{3}{c}{} & \multicolumn{3}{c}{Degree} & \multicolumn{1}{c}{}\\
        \cmidrule(lr){4-6}
        Graph & Nodes & Edges & Avg & Max & Min & Kind \\
        \hline
        \text{can\_1054}     & 1054 & 5571  & 10.57  & 34  & 5  & Structural\\
        \text{collins\_15NN} & 1000 & 8246  & 16.49  & 24  & 15 & Undirected Weighted\\ 
        \text{cnae9\_10NN}   & 1080 & 9139  & 16.92  & 221 & 10 & Undirected Weighted\\
        \text{delaunay\_n10} & 1024 & 3056  & 5.97   & 12  & 3  & Undirected\\
        \text{dwt\_1005}     & 1005 & 3808  & 7.58   & 26  & 3  & Structural\\
        \text{jagmesh2}      & 1009 & 2928  & 5.80   & 6   & 3  & 2D/3D\\
        \text{lshp1009}      & 1009 & 2928  & 5.80   & 6   & 3  & Duplicate Thermal\\
        \text{mice\_10NN}    & 1077 & 6742  & 12.52  & 22  & 10 & Undirected Weighted\\
        \text{msc01050}      & 1050 & 12574 & 23.95  & 127 & 4  & Structural\\
        \hline
    \end{tabular}
    \caption{Characteristics of graphs from the scale test dataset.}
    \label{tab:scale_tdata_info}
\end{table}

The second test set is used to test the neural acceleration framework at scale.
We use 9 graphs taken from the SuiteSparse Matrix Collection.
Information about these graphs is provided in Table~\ref{tab:scale_tdata_info}.
These graphs represent real-world problems such as structural, thermal, and 2D/3D, inheriting different structural properties.
By incorporating varying types of graphs, it enables us to evaluate the ability of our model to generalize to unseen data.
Additionally, the effects of different coarsening algorithms can be studied.

\subsection{Experimental Environment}
\label{subsec:exp-env}
The neural network model is developed using the PyTorch framework (version 2.2.2) in a Python 3 environment. The dataset is obtained through the SuiteSparse interface in MATLAB 2022a.
The system is powered by an Intel Core i7-10750H, featuring 6 physical cores and 12 threads, operating at a frequency of 2.6 GHz. 
The system is equipped with 24 GB of DDR4 RAM (1x8GB and 1x16GB at 2667 MHz).
The model is trained using the Nvidia GeForce GTX 1660 Ti GPU on the system with 6 GB of GDDR6.

\subsection{Network Training}
\label{subsec:net-training}
Our neural network model is trained in a supervised manner, i.e., expected inputs and outputs to minimize the L1Loss, also known as Mean Absolute Error (MAE) loss. The objective function is as follows:
\begin{equation}
    \min  \frac{1}{N} \sum_{i=1}^{N}  |\hat{y}_i - y_i|   
\end{equation}
where \begin{math} \hat{y}_i \end{math} is observed, and \begin{math} {y}_i \end{math} is the predicted output from the neural network model.

We use Adagrad~\cite{ada} as our optimizer to train our model. 
The learning rate ($\alpha = 0.001$) is explicitly set, while all other parameters were left at their default values as implemented in the PyTorch framework~\cite{pytorch}. 
Adagrad is chosen due to its adaptive learning rate mechanism, which dynamically adjusts the learning rate based on the accumulation of past gradients. 
This characteristic makes Adagrad particularly effective in handling sparse gradients and complex optimization surfaces, which are common in our problem domain.

We initially considered vanilla SGD~\cite{sgd} as our optimizer. 
However, we observed that SGD required more iterations to converge and frequently failed to reach an optimal solution within the given computational constraints. 
Since one of the main goals of our work was to achieve efficient training at compile time, the longer convergence time of SGD became a limiting factor. 
In contrast, Adagrad demonstrated faster convergence with fewer iterations, reaching an optimal solution more quickly. 
Although Adagrad has a trade-off in terms of increased memory consumption and additional floating point operations due to the need to store and accumulate squared gradients, the resulting performance gains in training efficiency outweighed these costs. 
This trade-off proves beneficial as it allowed us to achieve accurate model performance within a reasonable timeframe, which is critical for our use case.

\section{Experimental Evaluation}
In this section, we evaluate the use of our neural acceleration model for graph partitioning. 
This evaluation takes the form of quality and performance.
We consider the standard metrics of edge cut (equation~\ref{eq:edge-cut}) and edge ratio (equation~\ref{eq:edge-cut-ratio}) for quality.
We first provide measurements of quality for the train/test set.
We follow this with quality of the scaling test set.
Lastly, we present the performance metric in terms of timing and model size.
We provide results for our model's predicted partitions with and without refinement, referred to as NN-FM and NN, respectively.

\subsection{Evaluation of Sample Set Quality}
\label{subsec:eval-sample-set}

\begin{figure}[tbh]
    \centering
    \includegraphics[width=.4\textwidth]{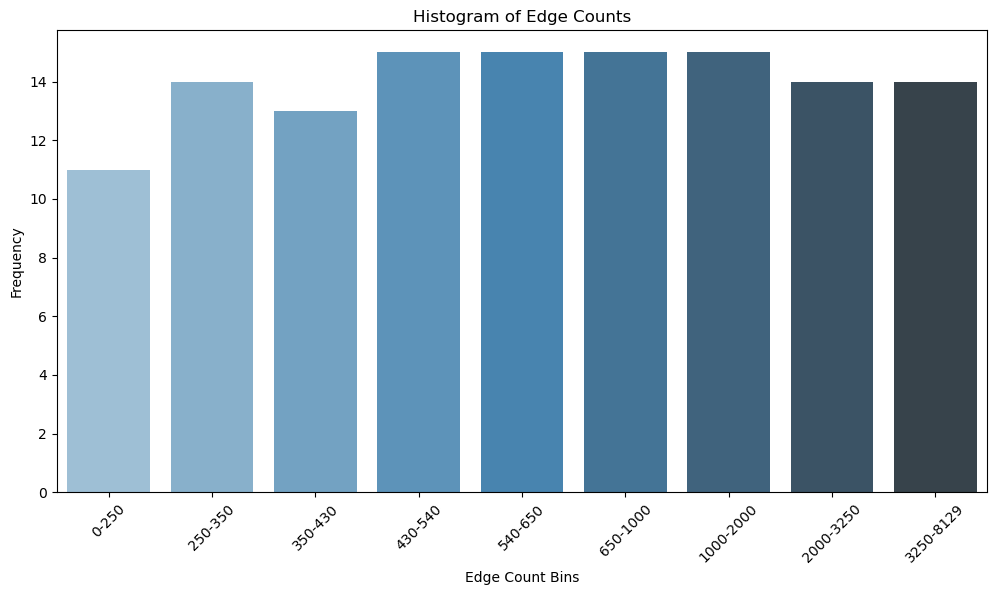}
    \caption{Grouped Graphs based on Number of Edges}
    \label{fig:bins-edge-count}
\end{figure}

In order to evaluate the overall quality of the model on all 126 test graphs, we consider the geometric mean (geomean) of the approximation ratio for each graph.
When evaluating approximate solutions against a certain problem, the approximation ratio is used to determine the quality and efficiency of the approximation algorithm.
The approximation ratio is simply the ratio of the edge cut of our predicted partition to the edge cut of the spectral partition. Formally, it is given as follows:
\begin{equation}
    approximation~ratio = \frac{EC_{NN}}{EC_{SPEC}}~.
\end{equation}
A ratio greater than 1 means worse performance, while the ratio less than or equal to 1 signifies better performance.
The lower the approximation ratio, the better the algorithm approximates the real problem.
\begin{figure}[tbh]
    \centering
    \includegraphics[width=.4\textwidth]{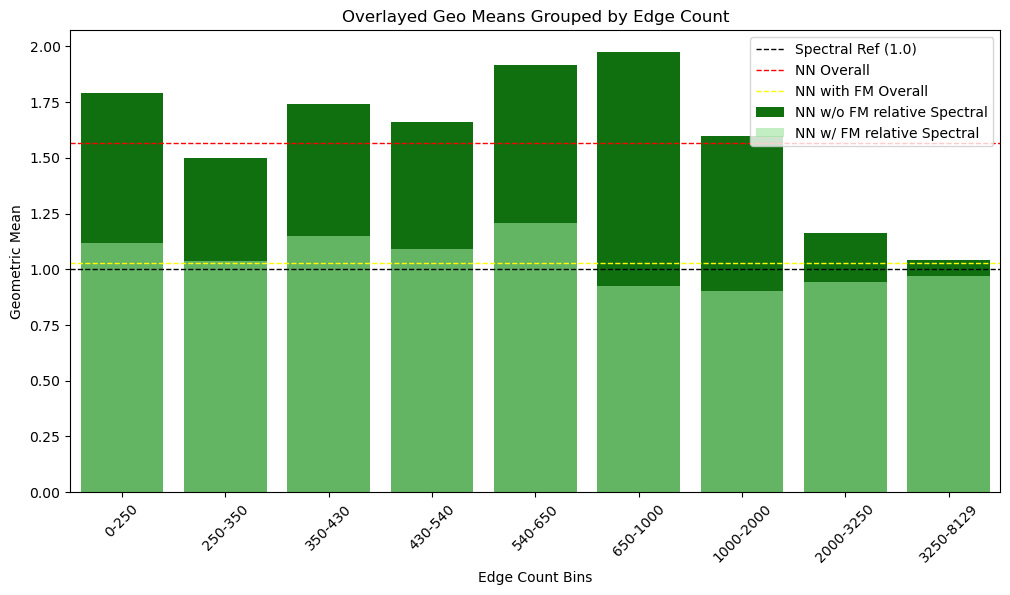}
    \caption{Overlap Geometric Means}
    \label{fig:geo_mean}
\end{figure}
The approximation ratios are calculated for NN and NN-FM.
To further explore the quality of our model based on the density of the test graphs, we group the test graphs based on their number of edges and provide the geomean of the approximation ratio for each group. 
Figure~\ref{fig:bins-edge-count} provides a histogram of the grouped graphs.
We overlap the geomean of NN and NN-FM for each to better understand the quality of our model.

\begin{table}[t]
\footnotesize
    \centering
    \begin{tabular}{ccccccc}
    \hline
        Graph & RAND & RAND-FM & METIS & SPEC & NN & NN-FM \\
        \hline
        4   & 216   & 62    & 11    & 15    & 63    & 18\\
        8   & 105   & 45    & 23    & 30    & 85    & 50\\
        19  & 419   & 97    & 76    & 106   & 138   & 86\\
        34  & 1327  & 1150  & 1130  & 1327  & 1324  & 1157\\
        45  & 226   & 47    & 26    & 31    & 42    & 36\\
        66  & 3595  & 3435  & 3433  & 3478  & 3590  & 3434\\
        88  & 69    & 25    & 1     & 1     & 1     & 1\\
        106 & 161   & 22    & 9     & 11    & 12    & 12\\
        109 & 177   & 58    & 20    & 20    & 22    & 22\\
        115 & 87    & 46    & 7     & 7     & 69    & 7\\
        \hline
    \end{tabular}
    \caption{Our Test Set Edge Cut Results}
    \label{tab:tdata_ec}
\end{table}

In Figure~\ref{fig:geo_mean}, it is evident that the NN-FM predictions are extremely close to the spectral than the NN.
NN-FM approximates the spectral partition within a factor of 1.03x, whereas NN is 1.56x worse.
The results of NN are expected as most of the graph partitioning algorithms perform refinement after an initial partition is generated, e.g., METIS.
NN provides a quick initial partition similar to spectral without the overhead of spectral methods, and allows refinement to produce edge cuts that are comparable to or even better than those from spectral.
We observe this trend with our results.
Furthermore, we observe that graphs with lower densities underperformed spectral by slightly more than the graphs with high densities.

In order to get a closer understanding, we next study the 10 random graphs from the test set.
Tables~\ref{tab:tdata_ec} and \ref{tab:tdata_ecr} provide edge cuts and edge cut ratios, respectively, along with random partitioning (RAND), spectral partitioning (SPEC), and METIS for comparison.
We note that RAND algorithm is simply assigning each vertex to a partition randomly, and that METIS already uses FM.
NN consistently approximates the spectral partition within a small factor, such as graphs \textit{45}, \textit{88}, \textit{106}, and \textit{109}, while outperforming RAND for the majority of the graphs.
NN and RAND produce similar results for graphs \textit{34} and \textit{66}, indicating the inherent difficulty of those instances.
\begin{table}[t]
\footnotesize
    \centering
    \begin{tabular}{cccccc}
    \hline
        Graph & RAND & RAND-FM & SPEC & NN & NN-FM  \\
        \hline
        4   & 0.511 & 0.147 & 0.035 & 0.149 & 0.043\\
        8   & 0.547 & 0.234 & 0.156 & 0.443 & 0.260\\
        19  & 0.503 & 0.116 & 0.127 & 0.166 & 0.103\\
        34  & 0.510 & 0.442 & 0.510 & 0.509 & 0.445\\
        45  & 0.538 & 0.112 & 0.074 & 0.100 & 0.086\\
        66  & 0.505 & 0.482 & 0.488 & 0.504 & 0.482\\
        88  & 0.543 & 0.197 & 0.008 & 0.008 & 0.008\\
        106 & 0.498 & 0.068 & 0.034 & 0.037 & 0.037\\
        109 & 0.488 & 0.160 & 0.055 & 0.061 & 0.061\\
        115 & 0.439 & 0.232 & 0.035 & 0.348 & 0.035\\
        \hline
    \end{tabular}
    \caption{Our Test Set Edge Cut Ratio Results}
    \label{tab:tdata_ecr}
\end{table}
NN-FM maintaines its advantage over RAND-FM, achieving significantly better results as seen in graphs \textit{4}, \textit{19}, and \textit{115}.
NN-FM demonstrates competitive performance against the SPEC method, comparable or better in some instances, including graphs \textit{88}, and \textit{115}.
NN's performance suggests it provides a strong alternative to traditional methods, especially in scenarios where FM refinement is applied.

\subsection{Evaluation of Scaling Set Quality}
\label{subsec:eval-scale-set}
Next, we evaluate the ability of our method to scale to large graphs in terms of quality.
We employ the multilevel approach to test the scalability or our model to partition large graphs.
The graphs are coarsened using our sliding window (SW) technique and HEM to further understand the effect of coarsening on partitioning algorithms.
The coarsest graphs produced by SW and HEM are different with varying number of edges.
We provide the results for both with the number of edges associated with each graph.

Table~\ref{tab:sw_hem_ec} provides edge cuts obtained by our model for both coarsening techniques.
Note that the results are provided at their coarsest level and no refinement is applied.
For the SW coarsening approach, NN provides remarkably close approximations to spectral.
It is seen in graphs, such as \textit{collins\_15NN}, \textit{delaunay\_n10}, and \textit{lshp1009}, that the edge cuts produced are only a few node refinements away.
These results are expected because the model was trained with graphs coarsened using SW.
It further indicates the model's generalization capability to larger graphs coarsened using the same technique.
And again, RAND still underperformed as expected.

\begin{figure}[tbh]
    \centering
    \centering
    \subfloat[Actual]{
        \includegraphics[width=.30\textwidth]{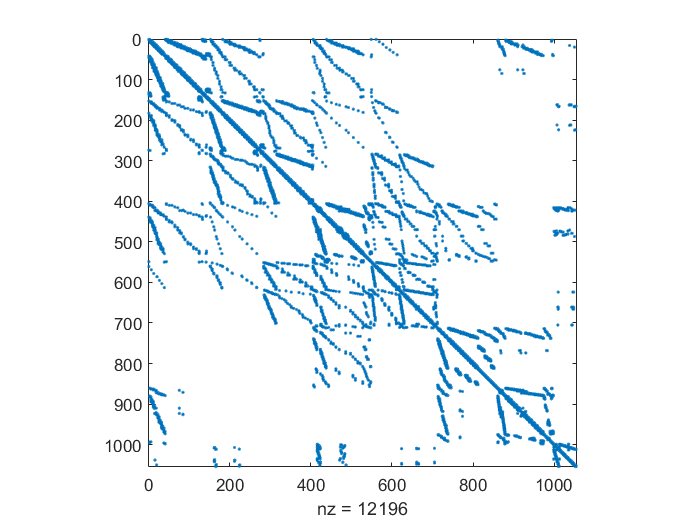}
    }
    \hfill
    \subfloat[SW]{
        \includegraphics[width=.30\textwidth]{./figures/can_1054_act_struct.png}
    }
    \hfill
    \subfloat[HEM]{
        \includegraphics[width=.30\textwidth]{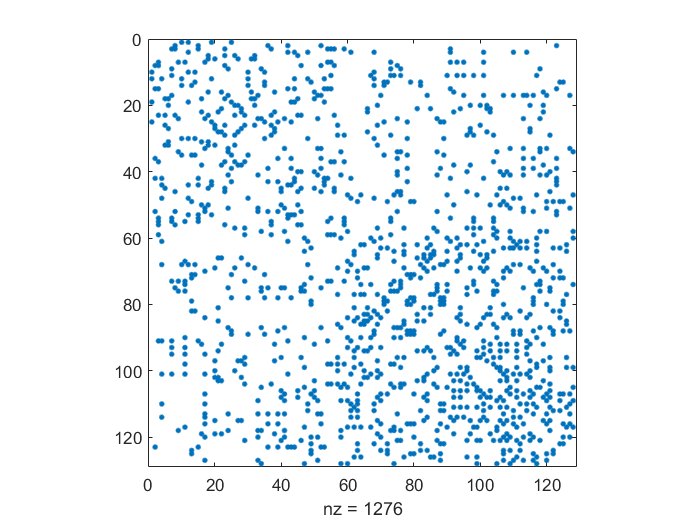}
    }
    \caption{\textit{can\_1054} Adjacency Matrix Structure}   
    \label{fig:adj-struct-eg}
\end{figure}

For the HEM coarsening approach, the edge cuts produced by our model significantly underperforms spectral, and slightly better than RAND.
NN produces better initial partition for SW partition than HEM, despite being the same graphs, is concerning.
To account for that behavior, we further investigates the structural difference between HEM and SW at the coarsest level by observing the shape of the adjacency matrices.
Figure~\ref{fig:adj-struct-eg} provides a visual of graph \textit{can\_1054} to analyze the effect of coarsening techniques.
You can notice that SW preserves the original structure extremely well, however, HEM displays a random distribution of values across the whole matrix.
Based on these observations, it is evident that our neural network learns the representation of the graphs much better and partitions the graph efficiently.
The argument of using HEM to coarsen the train set could arise, but the random distribution of values provides no patterns for the model to learn, making it difficult to train.
For such a reason, HEM works well with METIS as it performs multiple combinations of partitions at the coarsest level and picks the best.

\begin{table*}[tbh]
\footnotesize
    \centering
    \begin{tabular}{ c >{\centering\arraybackslash}p{1cm} >{\centering\arraybackslash}p{1cm} >{\centering\arraybackslash}p{1cm}>{\centering\arraybackslash}p{1cm} >{\centering\arraybackslash}p{1cm} >{\centering\arraybackslash}p{1cm}>{\centering\arraybackslash}p{1cm} >{\centering\arraybackslash}p{1cm}}
    \hline
        \multicolumn{1}{c}{} & \multicolumn{4}{c}{SW} & \multicolumn{4}{c}{HEM}\\
        \cmidrule(lr){2-5}
        \cmidrule(lr){6-9}
        Graph & Edges & RAND & SPEC & NN & Edges & RAND & SPEC & NN\\
        \hline
        \text{can\_1054}        & 1004 & 501   & 235   & 269   & 638  & 323  & 57  & 274 \\
        \text{collins\_15NN}    & 210  & 110   & 2     & 5     & 373  & 171  & 3   & 181 \\
        \text{cnae9\_10NN}      & 5358 & 2722  & 2592  & 2685  & 1749 & 897  & 457 & 863 \\
        \text{delaunay\_n10}    & 658  & 341   & 81    & 88    & 371  & 185  & 29  & 192 \\
        \text{dwt\_1005}        & 474  & 238   & 53    & 98    & 496  & 262  & 57  & 205 \\
        \text{jagmesh2}         & 425  & 225   & 38    & 44    & 358  & 183  & 23  & 174 \\
        \text{lshp1009}         & 425  & 219   & 38    & 44    & 358  & 194  & 23  & 174 \\
        \text{mice\_10NN}       & 560  & 281   & 67    & 83    & 670  & 339  & 58  & 308 \\
        \text{msc01050}         & 966  & 487   & 178   & 283   & 1434 & 730  & 471 & 604 \\
        \hline
    \end{tabular}
    \caption{External Test Set with SW and HEM Coarsening Edge Cut Results without FM Refinement}
    \label{tab:sw_hem_ec}
\end{table*}

\subsection{Performance}
\label{subsec:eval-perf}

\begin{figure}[tbh]
    \centering
    \subfloat[CPU]{
        \includegraphics[width=.4\textwidth]{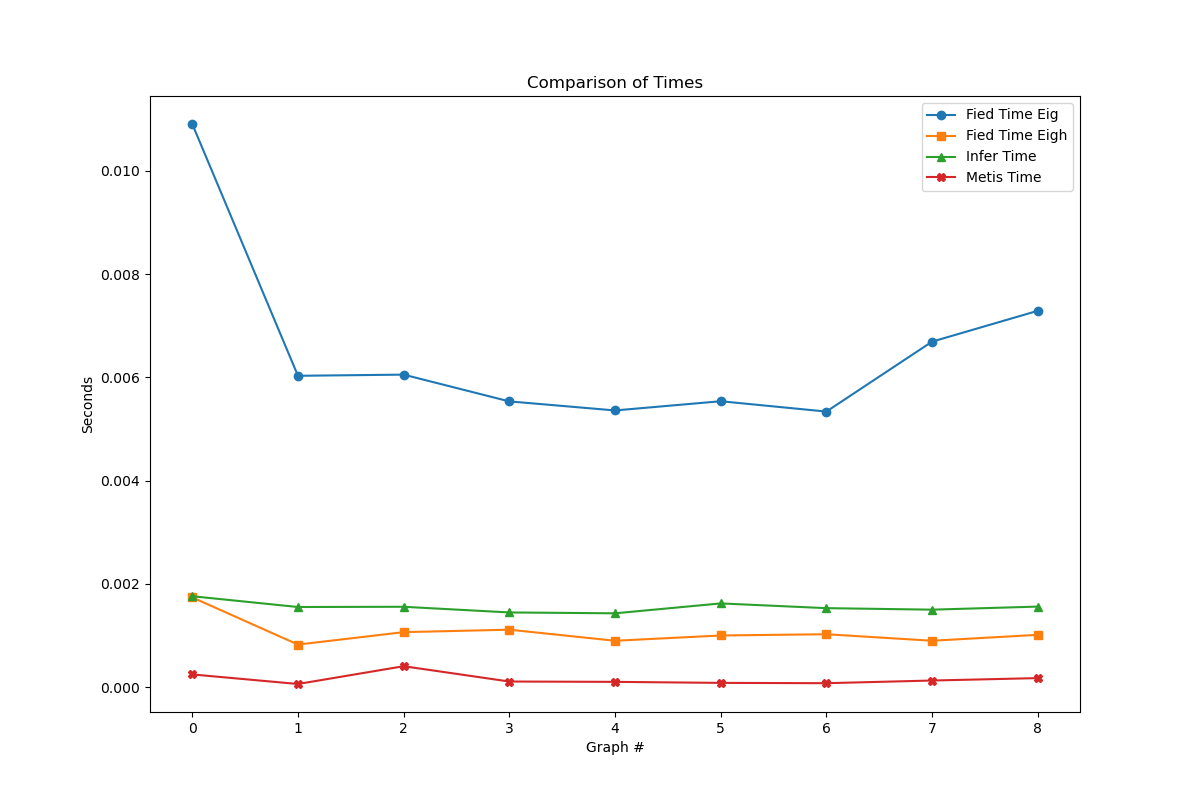}
    }
    \hfill
    \subfloat[GPU]{
        \includegraphics[width=.4\textwidth]{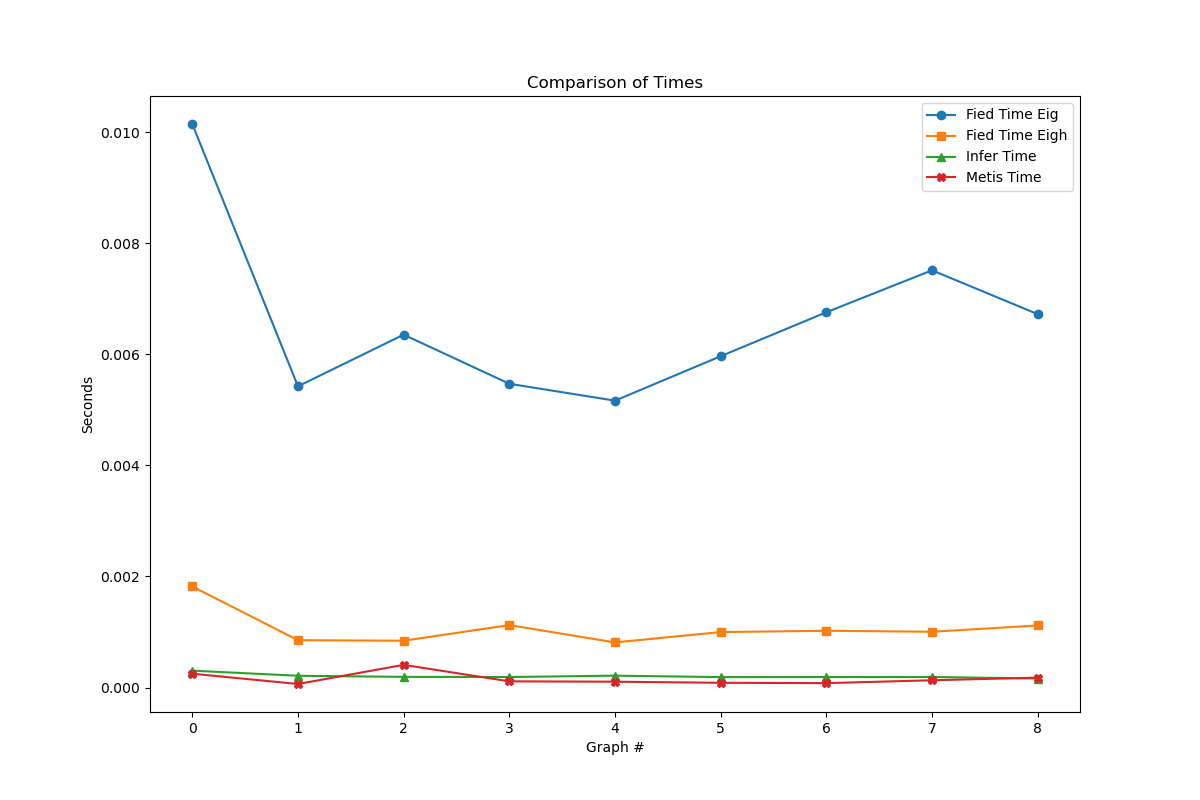}
    }
    \caption{External Test Set Execution Times}   
    \label{fig:ext-time}
\end{figure}

\textbf{Time Evaluation.} 
Next, we consider performance.
The critical timing aspect of our neural acceleration method is the model's inference time to predict the Fiedler vector.
Inference time was recorded on both CPU and GPU.
To compare our model's performance, we consider NumPy's Linear Algebra (linalg)~\cite{numpy} implementation of eig and eigh, and METIS.
These algorithms are timed on the CPU as the implementations cannot run on the GPU.
\begin{figure}[tbh]
    \centering
    \subfloat[CPU]{
        \includegraphics[width=.4\textwidth]{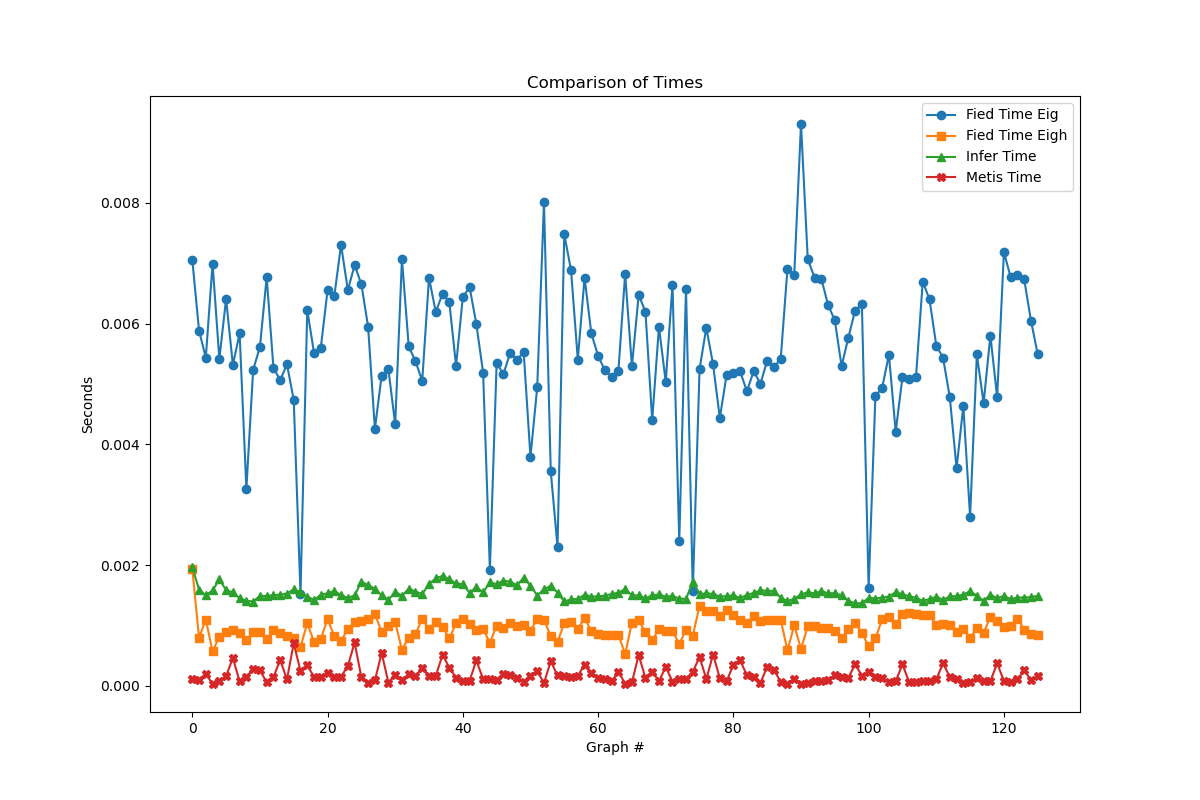}
    }
    \hfill
    \subfloat[GPU]{
        \includegraphics[width=.4\textwidth]{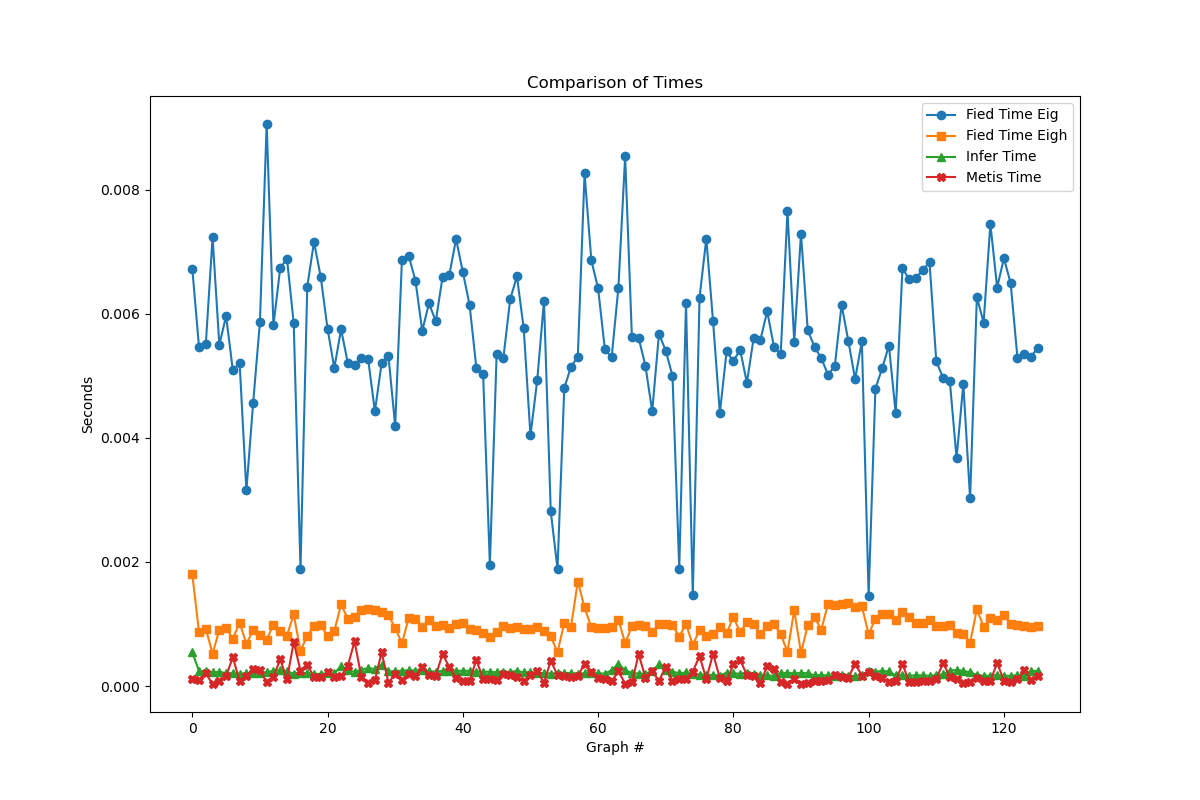}
    }
    \caption{Our Test Set Execution Times}   
    \label{fig:our-time}
\end{figure}
Note that inference, eig, and eigh times are recorded in an interpreted language (Python), and METIS is recorded using a complied language (C/C++).
The performance of interpreted languages tends to be worse than that of compiled languages.
It is important to consider the variability caused by the difference as we look at the timings.
We do not consider performance for the complete framework (as described in Figure~\ref{fig:EvalWorkflow}) because the main contribution of our work is the neural acceleration model.
Our SW coarsening technique can be implemented efficiently in a sparse format or in parallel to gain better performance. However, that is out of the scope of this work.
The timings of the sample and the scale test set are provided in Figures~\ref{fig:our-time} and \ref{fig:ext-time}, respectively.

In Figure~\ref{fig:our-time}, eigh and METIS outperform our model on the CPU by 1.6x and 8.3x, respectively.
The same trend is observed in Figure~\ref{fig:ext-time}.
We expected METIS to be faster than our model on the CPU due to its efficient implementation.
In contrast, our model performs 4.5x better than eigh, equivalent to METIS on the GPU.
We attribute these results to the GPU's optimization for efficient dense tensor multiplications.
Despite our model's high number of parameters, we show that it is as efficient or better than METIS and eigenvalue iterative solvers.

\begin{table}[t]
\footnotesize
    \centering
    \begin{tabular}{ c c c c c c}
        \hline
        Dataset & EIG & EIGH & METIS & NN-CPU & NN-GPU\\
        \hline
        Our & 0.698 & 0.121 & 0.023 & 0.193 & 0.027\\
        Scale & 0.0587 & 0.0096 & 0.0014 & 0.0139 & 0.0018\\
        \hline
    \end{tabular}
    \caption{Our and Scale Test Set Timings}
    \label{tab:timining}
\end{table}

\textbf{Size Evaluation.} Now, we address the issue of memory with respect to the number of parameters in our model. 
The total number of 8,454,656 trainable parameters amounts to roughly 34 MB of memory.
GPUs on recent systems, including personal computers, are equipped with 30-100x more memory than required to store our model and perform operations efficiently.
Therefore, our model achieves an execution time comparable to state-of-the-art methods on one GPU and requires minimal memory.

\section{Conclusion}
\label{sec:concl}

In general, energy-efficient methods are required to approximate solutions to many real-world problems that require immense computational resources.
Graph partitioning is such a problem with various applications in different fields.
The cost of processing and analyzing graph data increases tremendously with traditional graph partitioning methods.
We propose a neural acceleration model to approximate the spectral partitioning with great accuracy and efficiency.
Our method replaces the expensive calculation of the Fiedler vector by a simple dense neural network, which can be trained at compile time on a mid-range GPU within a few minutes.
Compared to spectral partitioning, we demonstrate that our method provides initial partition within a factor of 1.56x, and 1.03x with partition refinement.
In terms of execution time, the proposed method achieves a performance of 4.5x faster than traditional iterative solvers to perform eigen decomposition of a real symmetric matrix.
This method provides an alternative to graph partitioning methods.
Based on these results, it is conclusive that more applications can benefit by applying neural acceleration.

\bibliographystyle{acm}
\bibliography{main}

\end{document}